# Study of potential Hamiltonians for quantum graphity

Y. Spector and M. Schwartz

School of Physics and Astronomy, Raymond and Beverley Faculty of exact sciences,

Tel Aviv University, Tel Aviv 69978, Israel .

In this work we show that the simple Hamiltonians used in Quantum Graphity models are highly degenerate, having multiple ground states that are not lattices. In order to assess the distance of the resulting graphs from a lattice graph, we propose a new measure of the equivalence of vertices in the graph. We then propose a Hamiltonian that has a rectangular lattice as a ground state that appears to be non-degenerate. The resulting graphs are close to being a rectangular lattice, and the defects from the perfect lattice seem to behave like particles of quantized mass that attract one another.

**I.    Introduction**

It is long and well-known that while most forces in nature can be described within quantum mechanics, the gravitational force can only be described by the classical theory of general relativity and has at present no quantum mechanical description. The effort to combine the two has led researchers to attempt to formulate a theory of quantum gravity (QG). However, many problems arise when dealing with gravity on small length scales due to the assumption of the continuity of space, yielding unbounded results and singularities on small scales [1].

All of these problems would have been solved if one abandoned the assumption of space-time continuity and made it fundamentally discrete at the Planck scale. Such an approach would lose some of the symmetries of space and a good test for the theory would be if it was able to reproduce these symmetries and the classical results of gravity on large length scales. These symmetries and the classical geometry of space would then be an emergent phenomenon that should be evident in the classical limit.

Several approaches were proposed for the description of such a discrete space, among them loop quantum gravity [2], Causal dynamical triangulations [3], Causal sets, [4] and Quantum Graphity [5-10] to name a few. Our method is based on the Quantum Graphity model suggested by Konopka et al. in [5]. In this approach, no a priori geometry of space is assumed and the basic structure from which the geometry of space emerges in the classical limit is a graph. The graph is a mathematical formulation of discrete space and the geometrical properties including gravity are to be derived from it. Clearly, if that was the whole problem, a lattice like graph could have been postulated as a basic structure that corresponds in the classical limit to continuous flat space. Then any physical theory had to be formulated on that graph. The purpose of Quantum Graphity is more ambitious. It aims at starting from a most primitive graph and evolve it in time under certain prescribed rules in such a way that it will eventually be reminiscent, in the classical limit of the observed universe. This process is governed stochastically by a Hamiltonian. The Hamiltonian's eigenstates are graphs and the stochastic dynamics changes one state into another according to statistical mechanics principles. This Hamiltonian, assigns to a graph an "energy" which is expressed in terms of the adjacency matrix corresponding to the graph. Because the "energy" has to be independent on the specific representation of the graph in terms of an adjacency matrix it must be symmetric under permutation of the vertices of the graph.

It is reasonable to expect that the ground state of the Hamiltonian will correspond to empty space. Points in empty space are indistinguishable from one another and thus it stands to reason that the vertices in the ground state graph should be equivalent to one another. The equivalence of vertices seems to be a necessary condition for the ground state. There are many graphs, however, with equivalent vertices. In order that the ground state graph should correspond to flat $d$ dimensional space, we require that it must be an ordered lattice. Thus, we also expect the ground state to have the same number of nearest neighbors as a lattice with the same dimensionality as the desired emergent space. We also expect the low energy graphs to correspond to the present observable universe and thus to have consistent non-local macroscopic definitions of directions, distances etc.

In previous work [6, 7] the above ideas are used, for example, to construct valence and cycles-preferring Hamiltonians. Namely, those Hamiltonians prefer that the number of edges connecting each vertex to its nearest neighbors (the valence) is identical for all vertices and prefer the existence of cycles of a certain length. Both ref. [6, 7] show that while certain lattice-graphs are metastable states with local minima of energy for these Hamiltonians, they are not ground states and were not producible in Metropolis simulations. Wilkinson and Greentree (in [7, 8]) show that this basic model favors small disjoint subgraphs, and add a term preferring a desired number of $2^{nd}$-closest neighbors and $3^{rd}$-closest neighbors in order to keep the graph connected.

In this work, we use similar ideas and construct Hamiltonians that should prefer close-to-lattice structures. We consider connected non-directed graphs. Each such graph, $A$, is taken as an eigenstate of some Hamiltonian, $H$ with "energy" $H(A)$. We define the equilibrium probability to find the state $A$ by

$$P(A) = \exp(-\beta\, H(A)) / Z_G, \qquad (1)$$

where $\beta$ governs the degree that the statistical system prefers states close to the ground state, as the inverse of the temperature does when considering thermal equilibrium in statistical mechanics, and the canonical graph partition function is

$$Z_G = \sum_B \exp(-\beta\, H(B)), \qquad (2)$$

where the sum over $B$ is over all connected graphs.

We employ Metropolis dynamics governed by the equilibrium probability above to generate representative states and study their properties. An important question is what are the graphs that can be reached using a given model and what Hamiltonians should be used in order to reach a state that behaves correctly in the classical limit. In this work we attempt to answer this question by simulating the equilibrium states of Hamiltonians and finding their ground states.

We attempt to find simple Hamiltonians with a non-degenerate ground state that is a lattice. We study Hamiltonians that allow us to ensure that a lattice-like ground state does exist and assess how far a certain state is from that ground state. Finally, we show simulation results of various graph sizes (number of vertices) using these Hamiltonians and the structure that we reach, which are close to being a lattice.

The paper is organized as follows:

In the next section, we describe our QG model, entropy and energy, the possible interactions and the simulation methods employed. In section III, we consider a number of potential Hamiltonians and discuss their low energy states using zero temperature Metropolis simulations. In order to quantify the departure of a graph from a graph with equivalent vertices, which is necessary for flat space, we introduce an appropriate vertex equivalence measure. The Hamiltonians employed are constructed is such a way that their ground state energy is known and that they have a lattice-like ground state. We then discuss the properties of states close to the ground state in Section IV. Specifically we discuss the formation of defects in the rectangular lattice ground state. These defects can be identified as massive particles interacting via short range attraction. Lastly, the results are summarized in Section V.

**II. Energy and Entropy**

The basic premise of Quantum Graphity is that the primitive structure from which space emerges in the classical limit is a graph. A graph $G = \{V, E\}$ is composed of a set of $N$ vertices, $V$, and a set of edges between vertices in $V$, $E = \{(V_1, V_2) | V_1, V_2 \in V\}$. The graph is assumed to be undirected and without self-loops, and so every edge can be thought of as a subset of $V$ with 2 elements. So, each graph can be represented by an adjacency matrix $\mathbf{A}$, where $\mathbf{A}_{IJ} = 1 \Leftrightarrow \{(i.j) \in E\}$ and $\mathbf{A}_{IJ} = 0$ otherwise. This matrix is symmetric with a zero diagonal.

The Hamiltonian is expressed in terms of the elements of the adjacency matrix $\mathbf{A}$ defined above. The dynamical variables of the Hamiltonian are thus the different matrix elements in the upper triangle of $\mathbf{A}$. The Metropolis dynamics is affected by steps in which the value of the two matrix elements (i,j) and (j,i) may be switched from zero to one or from one to zero. Namely, in each step a single degree of freedom may change its value randomly according to the weights assigned by the Metropolis procedure. This implies that the equilibrium representative adjacency matrices obtained by the Metropolis scheme are actually governed by the probability to find the adjacency matrix **A**,

$$P(\mathbf{A}) = \exp(-\beta\, H(\mathbf{A})) / Z_\mathbf{A}, \qquad (3)$$

where the partition function of the adjacency matrices is

$$Z_\mathbf{A} = \sum_\mathbf{B} \exp(-\beta\, H(\mathbf{B})), \qquad (4)$$

the sum being over all adjacency matrices corresponding to connected graphs.

The "energy" associated with a specific adjacency matrix must depend on it in such a way that it will be invariant under permutations (relabeling of the vertices). It is easy to construct examples where this is not the case. A simple example of a Hamiltonian that is invariant under relabeling of the vertices is,

$$H(\mathbf{A}) = tr(\mathbf{A}^M) \qquad (5)$$

This Hamiltonian counts the number of loops of length $M$ in the graph. (Such a loop is any trajectory of length $M$ that begins and ends at the same vertex, even if it is self-repeating. Each such loop is counted actually $M$ times.) Thus it is directly related to the structure of the graph and not to its specific representation by an adjacency matrix. Therefore, such a Hamiltonian can also be written as a Hamiltonian of the graph (as opposed to the matrix) $H(A)$.

Once we have discussed the "energy" associated with a graph we turn to the concept of entropy of a graph and discuss its effect on our Metropolis procedure. To each graph $A$ correspond $N(A)$ adjacency matrices. We can write $N(A)$ also as $N(\mathbf{A})$, the number of adjacency matrices corresponding to the same graph as the adjacency matrix $\mathbf{A}$. The entropy of a graph will thus be defined naturally as

$$S(A) = \ln\{N(A)\}. \qquad (6)$$

A simple relation exists between, $Z_G$ the partition function of the unlabeled connected graphs, and $Z_A$ the partition function of the corresponding adjacency matrices.

It is clear that

$$Z_G = \sum_{\mathbf{B}} N^{-1}(\mathbf{B}) \exp(-\beta \, \mathrm{H}(\mathbf{B})) \qquad (7)$$

and that

$$P(\mathbf{A}) = \exp(-\beta \, \mathrm{H}(\mathbf{A})) / Z_G. \qquad (8)$$

Thus, in principle, the correct weight for the Metropolis dynamics when treating the elements of the adjacency matrix as dynamical variables is proportional to $\exp(-S(\mathbf{B}) - \beta \, \mathrm{H}(\mathbf{B}))$ rather than to $\exp(-\beta \, \mathrm{H}(\mathbf{B}))$. If $S(\mathbf{A})$ was independent of $\mathbf{A}$, not only the equilibrium properties but also the statistical dynamics would have been identical for the graphs and their representations by adjacency matrices following the Metropolis dynamics of adjacency matrices. The reader can convince himself by working out some simple examples that this is not the case and in fact $S(\mathbf{A})$ is not independent of $\mathbf{A}$.

It can be argued, however, that most connected graphs have *N!* different adjacency matrices corresponding to them (due to *N!* permutations of the labels of the vertices, *N* being the number of vertices in the graph). These graphs are called asymmetric graphs. A very small portion of the graphs have less than *N!* different adjacency matrices, because it may happen (though not very probably) that an adjacency matrix is invariant under certain permutations (The sub- group of permutation matrices that commute with the specific adjacency matrix.). These graphs are less probable in the labeled statistics (adjacency matrix statistics) than in the unlabeled statistics. In order for a graph to have less than *N!* different adjacency matrices corresponding to it, it has to be composed of separate sub-graphs where an operation such as rotation or inversion of labels with respect to an axis of symmetry on one sub-graph (a) does not affect the other sub-graphs and clearly (b) the corresponding adjacency matrix for that sub graph does not change itself under that operation. Illustrations of the above are given in Fig. 1. The larger *N* becomes, the smaller becomes the relative abundance of these graphs and even more so if we consider graphs which are connected and close to lattices. The exact number of asymmetric graphs is known for small *N* [22, 23]. We can then calculate and see the small *N* behavior of the ratio of the number of asymmetric graphs to that of all graphs in both labeled and unlabeled statistics (see Fig. 2). These calculations indicate that an overwhelming majority of graphs is asymmetric even for small *N*, and it is expected to approach unity for large *N*.

Thus, the statistical properties of the system are set primarily by the more abundant asymmetric graphs which all have the same number of different permutations. Thus, the labeled graph statistics is an excellent approximation of the unlabeled graph statistics and we can safely use the Metropolis method, with the elements of the adjacency matrix as dynamical variables.

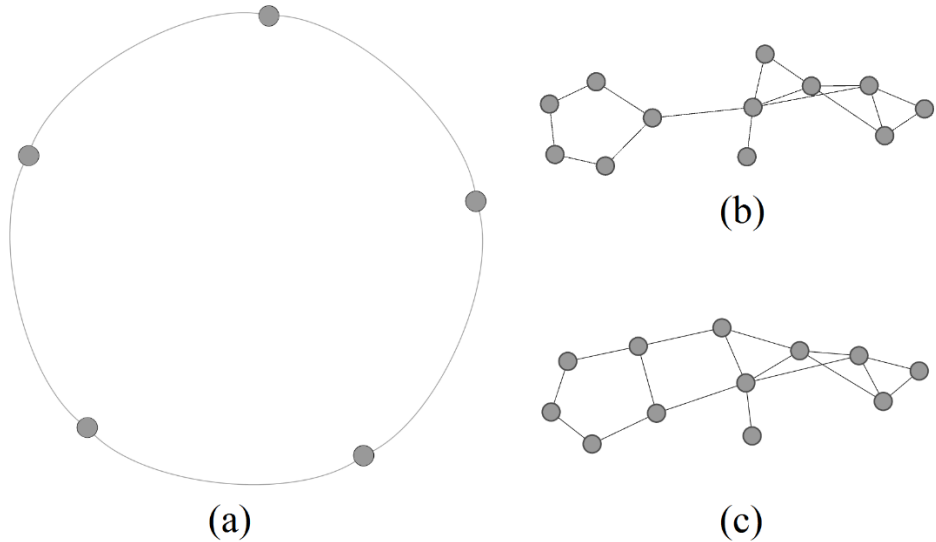

(a)          (b)          (c)

FIG. 1. Symmetries in subgraphs such as rings and fully connected mesh structures will retain their invariance to permutations only if they are connected to the rest of the graph by a single axis of symmetry with respect to these permutations. (a) A ring graph, having *N*=5 and 12 different possible adjacency matrices out of the 120 possible permutations. (b) A graph built from a ring graph on a "thread" connecting it to the rest of the graph. The ring subgraph retains its symmetry to permutations as long as the connecting vertex remains in place. (c) The same graph as in (b) with an additional edge no longer retains its symmetry to permutations.

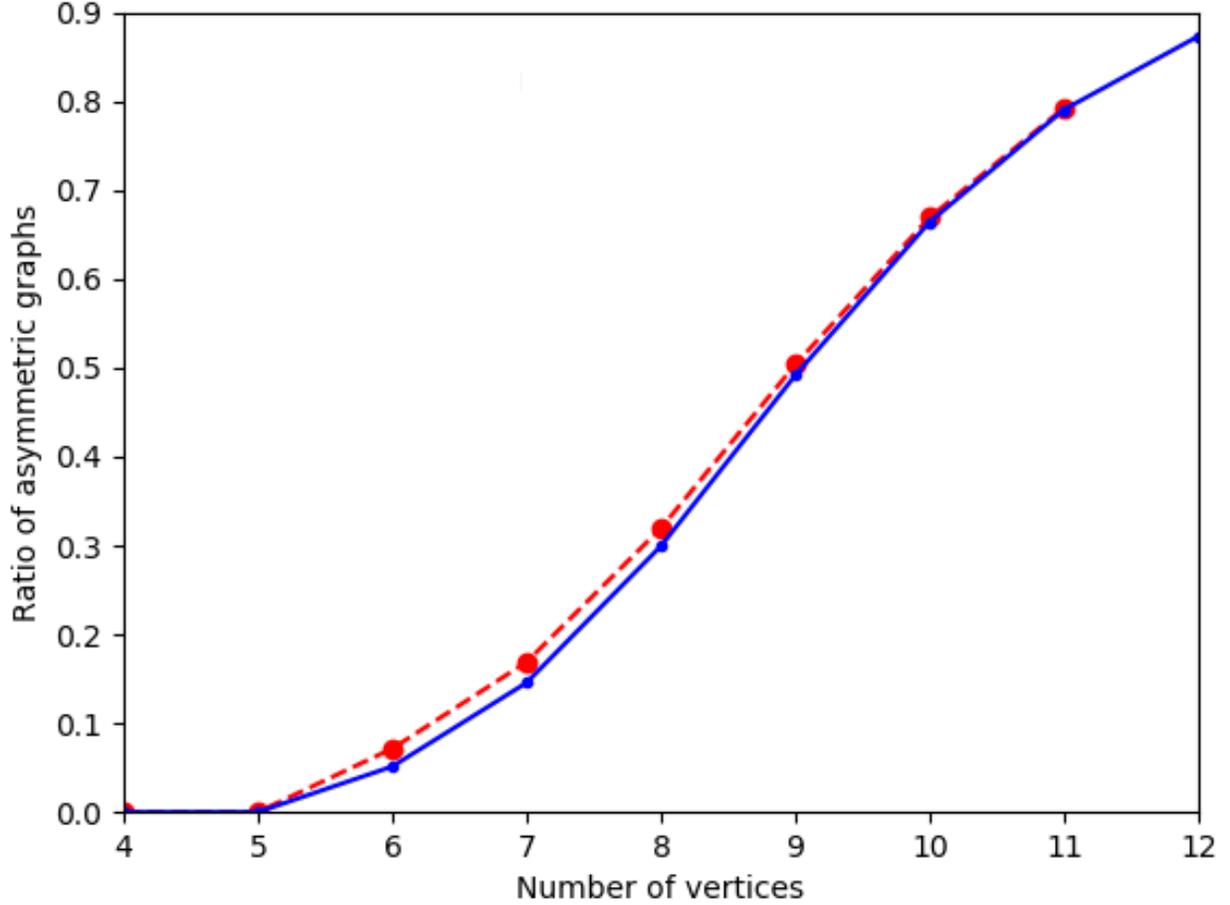

FIG. 2. Ratio of unlabeled asymmetric graphs to all unlabeled graphs of size *N* for small *N*: continuous (blue) line. The same ratio for connected unlabeled graphs: dashed (red) line.

### III. Hamiltonians

#### A. Valence term

The most basic Hamiltonian we use is a Hamiltonian with a single valence term.

The number of neighbors of a vertex is called that vertex's valence. It is also the number of edges connected to that vertex which is called the vertex's degree. A vertex's valence is the sum of the corresponding row (or column) in the adjacency matrix:

$$V_j = \sum_i A_{ij} \, . \tag{9}$$

For hyper cubic graphs, the number of neighbors in the lattice equals to 2 times the lattice dimension. As this means that vertices' valence is, at least intuitively, connected to the number of dimensions in a lattice-like space, we use a term

that will prefer a specific valence $Z$ for every vertex in the graph:

$$H_V = g_V \sum_i \left(\sum_j A_{ij} - Z\right)^2 . \qquad (10)$$

This term is minimal (with 0 value) when every vertex has valence $Z$.

If we do not impose connectivity of the graph, the Metropolis simulation at low temperature yields multiple different ground states with zero "energy". For the special case, $Z=2$, the resulting representative ground state graphs consist of multiple disconnected rings of various sizes, in agreement with the results of Wilkinson and Greentree [7, 8].

When $Z>2$, the resulting ground states are disordered tree like graphs but are usually connected, as opposed to the case of $Z=2$.

The rings in the $Z=2$ case are actually one-dimensional (1d) lattices with periodic boundary conditions. In a 1d universe, we would expect flat space to correspond to a single 1d lattice (single ring) which is indeed the ground state graph obtained when we impose connectivity.

### B. Imposing connectivity

We could think of imposing connectivity by adding a term to the Hamiltonian that would prefer connected graphs over graphs which are not connected. Such a term could be,

$$H_C = g_C (N_c - 1)^2 , \qquad (11)$$

where $N_c$ is the number of components (connected sub-graphs that are mutually disconnected). An extremely large coupling constant, $g_C$, will ensure a high "energy" penalty for breaking up a connected graph. In practice, we impose the constraint as a hard condition by modifying the Metropolis procedure. We start from a connected graph and during the procedure, we check if a suggested move of removing an edge breaks that graph into two disconnected sub-graphs. If it does, that move is aborted, regardless of any energy considerations.

The resulting ground state, for $Z=2$, is as expected: a single ring connecting all vertices. It can be proven that this is the only state that minimizes the energy of the Hamiltonian, as this is the only state where every vertex has valence 2 and it consists of a single connected component. It is worth noting that while the Metropolis yields different rings in every simulation run in terms of the labeled graph, they all correspond to the same unlabeled graph.

Attempting this with $Z>2$, however, does not lead to lattice-like results. While the simulation converges to a ground state almost always, there are numerous (unlabeled) ground states and it appears that most of them are disordered, having different properties at different vertices (see for example Fig. 3-5). These multiple ground states belong to the family called random-regular graphs. We offer here a number of local properties that distinguish among vertices in a graph and some global properties of graphs that distinguish among graphs. We are going to use these in the following and show that indeed all the ground states we present are quite different.

We define; (a) Number of triangles associated with the vertex i: The number $N_T(i)$, of different (i.e. having a different set of vertices) closed loops from the vertex i to itself of length 3.

(b) The distance $d_{ij}$, between the vertex i and the vertex j is the minimal number of edges needed to go from i to j.

(c) Eccentricity of the vertex i is the distance from the vertex i to the vertex furthest from it.

$$\epsilon_i = \max\{d_{ij}\}, \qquad (12)$$

where the maximum is taken over j with i fixed.

(d) The total number of triangles in the graph, $N_T$.

(e) Average path length of the graph:

$$l_G \equiv \frac{1}{N(N-1)} \sum_{i \neq j} d_{ij}. \qquad (13)$$

(f) The graph diameter is the distance between the 2 furthest vertices in the graph.

$$D = max\{\epsilon_i\}. \qquad (14)$$

Our results are consistent for various valences Z and graph sizes N. In these graphs it is difficult to define spatial dimensions in any non-local way, even at low resolutions.

Using the above properties, we made sure that the resulting states do not include different adjacency matrices corresponding to the same graph.

### C. Vertex equivalence measure

As stated previously, we expect the required ground state to be a graph where all vertices are equivalent, which will always be the case in an ordered lattice with periodic boundary conditions. This motivates us to introduce a measure of deviation of a graph from having that necessary property. Note, that this is necessary but not sufficient to have a lattice like structure.

We propose a measure called shell equivalence distance - $D_{SE}$: for each vertex we calculate the number of nearest neighbors (size of 1$^{st}$ shell), second-nearest neighbors (2$^{nd}$ shell) and so on until we arrive at an empty shell. For each pair of vertices we calculate the sum of the square differences between each of their shell sizes. The result is then normalized by the number of pairs of vertices:

$$D_{SE} = \sum_{i \neq j} \sum_{k=1}^{N_s} \frac{[S_k(i) - S_k(j)]^2}{N(N-1)/2} , \quad (15)$$

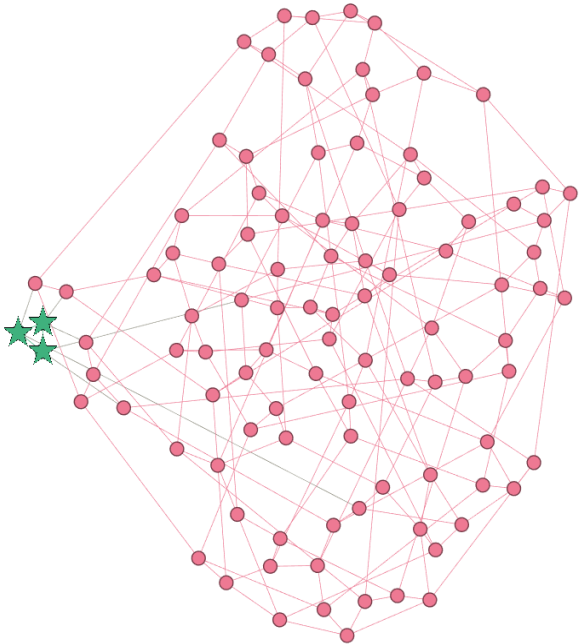

FIG. 3. Ground state result for $N$=100, $Z$=4 under the valence Hamiltonian H=$g_V H_V$ with imposed connectivity. Vertices in green stars are members of a single triangle and vertices in red circles are not members of any triangle. This graph has $N_T$ =1, D=6, $I_G$=3.528, $D_{SE}$=98.26.

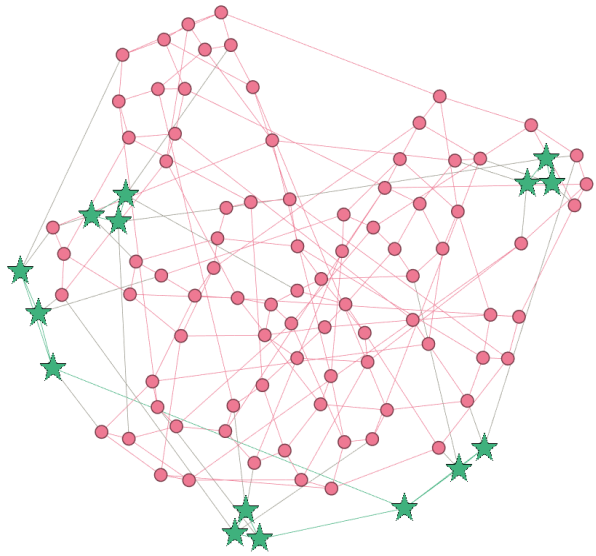

FIG. 4. Ground state result of the valence Hamiltonian $H=g_V H_V$ with imposed connectivity for $N$=100, $Z$=4. Green star vertices are members of a single triangle, red circle vertices are not members of any triangle. This graph has $N_T$=5, D=7, $I_G$=3.598, $D_{SE}$=105.47.

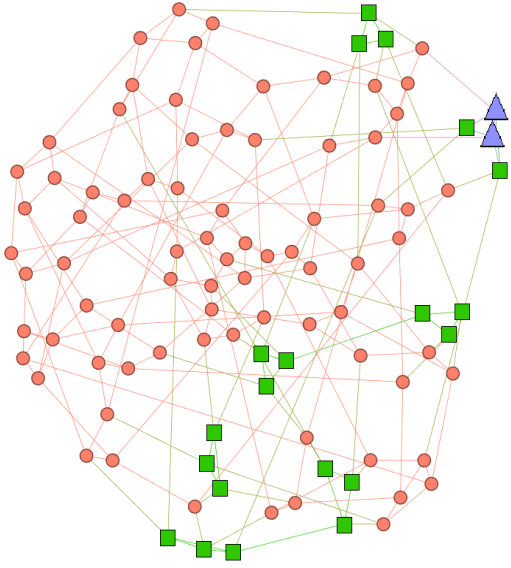

FIG. 5. Metropolis ground state result for the Hamiltonian H=$g_V H_V$ with imposed connectivity for N=100 and Z=4. Vertices in green squares are members of a single triangle while triangle vertices in blue are members of 2 triangles. The majority red circle vertices are not members of any triangle. The graph has $N_T$=8, D=6, $l_G$=3.583, $D_{SE}$=132.4.

where $N_s$ is the maximum number of non-empty shells that any vertex has in the graph and $S_k(i)$ is the size of the k-th shell of vertex i.

Obviously, for a perfect lattice the shell equivalence distance vanishes.

### D. Triangle sensitive term

In order to make the ground state more lattice-like, we add terms to the Hamiltonian that will prefer certain desired structures. In particular we chose to add a term that counts triangles in order to induce certain configurations:

We notice that the cubic lattice is a graph with Z=6 nearest neighbors for every vertex and with zero triangles, while the triangular lattice is also with Z=6 but each vertex is a member of 6 triangles. This means that the Hamiltonian

$$H_T = g_T \sum_{i=1}^{N} ((A^3)_{ii} - 2N_T)^2, \quad (16)$$

with $N_T = 0$ (meaning every vertex should be a member of zero triangles) has a ground state which is a cubic lattice (3-dimensional) with periodic boundary conditions, and the same Hamiltonian with $N_T = 6$ (meaning every vertex should be a member of 6 triangles) has a ground state which is a triangular lattice (2-dimensional) with periodic boundary conditions.

However, when we ran simulations with the Hamiltonian $H_V + H_T$ with Z=6 and $N_T = 0$, we arrived at multiple ground states that are not a cubic lattice and did not arrive at any cubic lattice ground state. We used the vertices' eccentricity to make sure the ground states are not equivalent. Some results can be seen in Fig. 6.

The same Hamiltonian ($H_V + H_T$) with $N_T = 6$ or a similar Hamiltonian that prefer triangles (with negative $g_T$ and $N_T = 0$) did not reach a ground state at all.

In general, the convergence of the simulation for these Hamiltonians that have higher power of the adjacency matrix was much slower and seldom arrived at the ground state.

Our conclusions from these simulation runs are that these high-power terms are not enough to reduce the degeneracy of the ground state, and that they affect convergence in a dramatic way. We want to find other terms that will reduce the degeneracy by being less tolerant to configurations that deviate from lattice structures, but that will be as local and with as low powers of the adjacency matrix as possible.

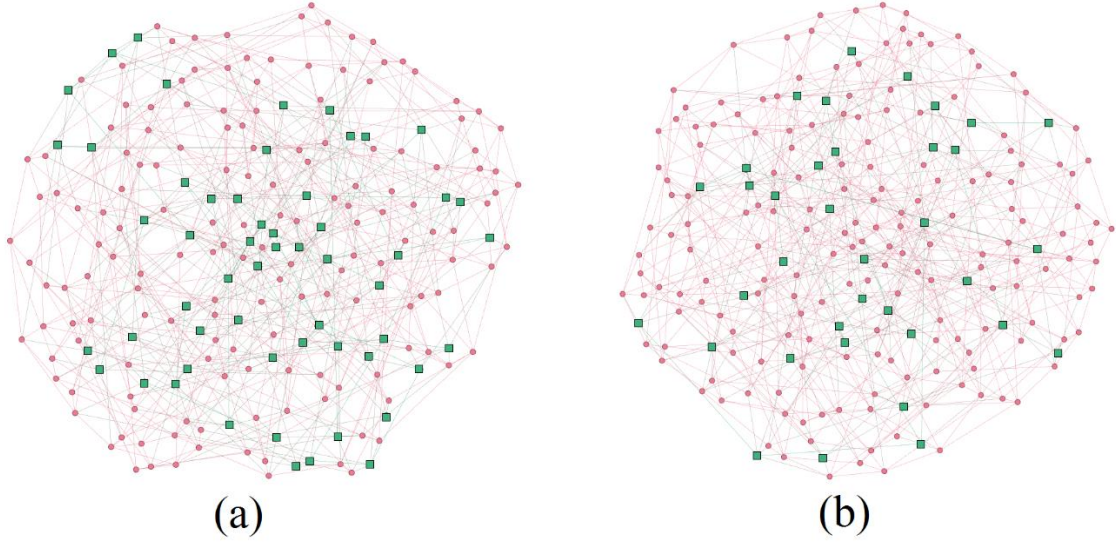

FIG. 6. Ground states of Hamiltonian that prefers zero triangles ($H = g_V H_V + H_T; Z = 6; N_T = 0; g_V/g_T = 50$) for N=216. The 2 graphs are not equivalent due to having different eccentricity frequencies. Red circle vertices have eccentricity $\varepsilon_i=5$ while green square vertices have eccentricity $\varepsilon_i=4$. In (a) The shell equivalence distance is 346.75. In (b) its value is 383.1.

### E. Directions generating term

In the following we construct a Hamiltonian that has a *d*-dimensional cubic lattice as a ground state, by preferring structures where *d* directions can be defined locally. We base our Hamiltonian on the observation that square lattices have *d* directions where each direction can be defined at a vertex by the 2 neighboring vertices that have a single path of length 2 between them (see Fig. 7).

The following term is minimal (and equals zero) when there are exactly 2·*d* couples of neighbors of every vertex that have a single path of length 2 between them, as would be in a hyper-cubic *d*-dimensional lattice with periodic boundary conditions:

$$H_D \propto \sum_i \left( \sum_{\substack{j,k \\ j \neq k}} \mathbf{A}_{ij} \mathbf{A}_{ik} \delta_{\mathbf{B}_{jk},1} - 2d \right)^2, \quad (17)$$

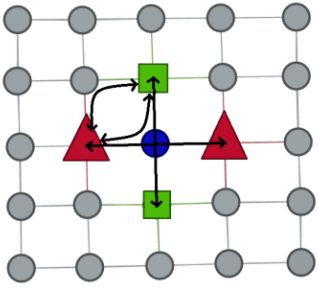

FIG. 7. A 2d square lattice showing our definitions of directions. The center vertex (Blue circle) has 2 horizontal neighbors (Red triangles) and 2 vertical neighbors (Green squares). The horizontal vertices have a single path of length 2 between them, as do the vertical ones. A vertical neighbor and a horizontal neighbor are connected by 2 paths of length 2.

where $\mathbf{B}=\mathbf{A}^2$. Using this term, the following Hamiltonian

$$H = H_V + H_D. \qquad (18)$$

has minimal "energy" for a cubic lattice of $d$ dimensions with periodic boundary conditions, when $Z=2d$.

We ran Metropolis simulations with the above Hamiltonian for $d=2$, and failed to reach a ground state (We monitored the "energy" and it never reached zero).

As we did not reach a ground state, we attempted to accelerate the simulation. We have done that by adopting a triple strategy. First we note that the adjacency matrix for a regular lattice is sparse and start from a ring graph rather than from the full graph. Then we note that for $Z>2$ the graphs obtained by the simulation are always connected. So, we drop the continuity condition. Finally we use a modified Metropolis method. Instead of choosing a random vertex and add or subtract a link connected to it according to the Metropolis weights, the simulation prefers choosing vertices where the next step is prone to lower the energy. In this way, the simulation does not "waste" moves on changes that are bound to raise the energy. This is similar to the epitaxial approximation in [8], for the Metropolis procedure.

We still did not reach a ground state, but came very close. Moreover, the graphs that we arrived at seemed close to being rectangular: Having relatively similar structure with very few defects (see Fig. 8). The minimum shell equivalence distance we found was 20.51 which is a lot smaller than $D_{SE}$ of all of our previous Hamiltonians.

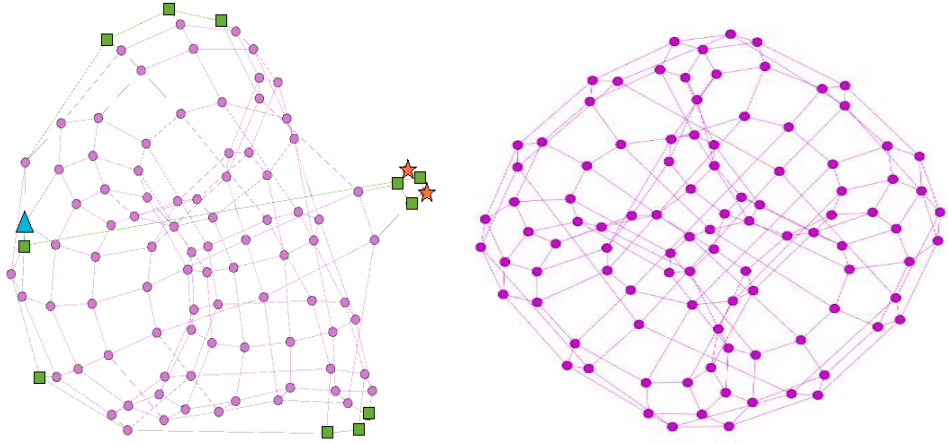

FIG. 8. Left: A metastable state of the Hamiltonian $H=g_V H_V + g_D H_D$ with $N$=100 ($g_V/g_D$=0.6). This graph has a shell equivalence distance of 20.51. Notice that this graph has many 4-loops and it seems that directions do exist in small neighborhoods even non-locally. The majority of the vertices (circles with pink color) have $Z$=4 neighbors. Green square vertices have $Z$=3 neighbors, red star vertices have $Z$=2 neighbors and the blue triangle vertex has $Z$=5 neighbors. Right: A perfect rectangular lattice with periodic boundary conditions and $N$=100. Both graphs are arranged using the same algorithm.

## IV. Defects

The Metropolis procedure with the directions generating Hamiltonian does not lead in reasonable time from the initial fully connected graph, or even a ring graph, to the required, square lattice, ground state. This raises an interesting question. What will happen if we start from the square lattice, disturb the system slightly and then let it relax. We used the Hamiltonian (eq. (18) with the connectivity constraint), at a finite temperature. Then we dropped the temperature back to zero. During the heating phase we observed the evolving states, relative to the initial ground state.

We found that the excited states obtained at low temperature could be characterized as defects in the ordered lattice structure. We will address the topics of defects later but first we discuss the outcome of the final fast cooling of the system.

### A. Non-degeneracy of the ground state?

Our simulations show that even with a small number of defects, when we lower the temperature the graph takes a very long time to converge back to the ground state, and it does not always do so in the normal Metropolis method. Our modified Metropolis method always converged back very fast, meaning that the excited states that do not return to the ground state by the ordinary Metropolis are not really metastable. The high entropy of possible moves makes it improbable for the simulation to randomly choose the exact defects that formed in the heating process in order to return to the ground state. When we do go back to a ground state, it is always the same ground state, even in terms of the adjacency matrix. This supports the conjecture that the ground state of this Hamiltonian is non-degenerate (i.e. there is only a single unlabeled ground state).

### B. Attraction of defects

Analyzing the defects formation during the heating phase, reveals that the energy of the graph depends on the number of defects, their type and their relative distance. We define 2 types of defects. Type A is the type of a missing edge relative to the square lattice and type B is an extra edge, relative to that lattice. The single defects have energies $\varepsilon_A$ and $\varepsilon_B$ respectively with $\varepsilon_B > \varepsilon_A$. The energy difference is substantial enough that an extra edge rarely forms at low temperatures. The ratio of the number of defects of type B to that of type A is given by

$$n_B/n_A = \exp[-\beta(\varepsilon_B - \varepsilon_B)]. \qquad (19)$$

For this reason, we deal here with type A defects only and postpone the analysis of type B to future publications. The defect breaks locally the symmetry between the horizontal and vertical directions. It can be expected that two defects will interact via a short range interaction dependent on the relative orientations of the missing edges in the two defects. By the tem "interaction" we mean, of course, that the energy of a pair of defects is not the sum of their individual energies. Indeed, when the defects are far away from one another, each missing edge contributes an equal amount of energy and the total energy is exactly proportional to the number of defects, leading us to think the defects may represent particles of quantized mass. However, when the two defects get closer, their energy decreases, leading us to think of these particles as having an attractive interaction between them (see Fig. 9).

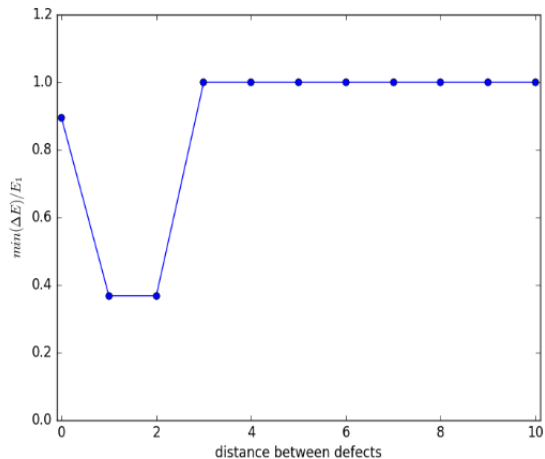

FIG. 9. Minimum over all relative orientations and positions of the interaction energy of two defects. This energy is presented as a fraction of the single defect energy. For short distances the needed energy is reduced, suggesting an attractive interaction between the defects.

## V. Summary and Conclusions

We have presented a number of potential Hamiltonians for Quantum Graphity. The various Hamiltonians were used to drive Metropolis stochastic evolution of graphs towards the ground states of those Hamiltonians. The ground states were chosen such that their low resolution limit would be a $d$-dimensional empty space. Thus the whole evolution is intended to describe the evolution of the universe from a pre-geometric phase into a phase where geometry exists. The dynamical degrees of freedom in the Metropolis procedure are the entries of the adjacency matrix. The first issue discussed is whether the statistics of graphs is indeed the statistics of adjacency matrices or not. We find that strictly speaking this is not the case but for large graphs it is not relevant. A second issue that had to be addressed is that of equivalence of vertices in a graph. Since, the ground state has the property that all the vertices are equivalent, we have introduced a new measure, which characterizes the departure of a graph from that property. It has been found that this measure is indeed useful in characterizing to what extent is a given graph close to the ordered lattice ground state.

It was easy, using the simplest Hamiltonian together with the connectivity constraint to reach a one dimensional lattice with periodic boundary conditions, from an initial state where all the vertices are connected. When working with the same Hamiltonian but characterized by a different parameters in such a way as to prefer a higher dimensional hyper cubic lattice with periodic boundary conditions, we have reached ground states but not the required lattice with periodic boundary conditions. Namely with the simplest Hamiltonian (the valence preferring Hamiltonian) it is easy to reach a ground state but the ground state is highly degenerate. The first conclusion is that in the end the valence Hamiltonian, although necessary, is not very useful in obtaining ordered lattices. Moreover, it seems that the same effect could be achieved by simulating under a hard constraint of a given number of neighbors. In order to lift the degeneracy, we used first a triangle counting interaction, which prefers ordered lattices. Starting from the fully connected graph, the Metropolis procedure, has not reached a ground state at all. Next we introduced a new interaction term that prefers local generation of $d$ directions. Although, we have not reached a lattice–like ground state, we have been able to reach within the time of the simulation a state much closer to a two dimensional square lattice than the states obtained with other interactions. Moreover, by starting from the lattice ground state heating up the system and then letting it cool down, we arrive at an indication that the unlabeled square lattice might be non-degenerate with the direction generating Hamiltonian. Study of excited states reveals defects that seem to have a given mass and interact via a short range interaction.

We believe that future work in this field will benefit from the shell distance measure and from direction generating interactions.